\documentclass[aps]{revtex4b3}
\usepackage{epsf,multicol}  

\begin{document}

\title{\Large \bf Dynamic Approach to Weak First Order Phase Transitions}

\author{\bf L. Sch\"ulke$^1$  and  B. Zheng$^{2,3}$}

\address{$^1$FB Physik, Universit\"at Siegen, D--57068 Siegen, Germany}
\address{$^2$FB Physik, Universit\"at Halle, D--06099 Halle, Germany}
\address{$^3$Institute of Theoretical Physics, Academy of Science,
            100080 Beijing, P.R. China}

\begin{abstract}
A short-time dynamic approach 
to weak first order phase transitions
is proposed. Taking the 2-dimensional 
 Potts models as examples,
from short-time behaviour of
non-equilibrium relaxational processes starting from
 high temperature and zero temperature states,x
pseudo critical points $K^{*}$ and $K^{**}$
are determined. A clear difference of 
the values for $K^{*}$ and $K^{**}$ distinguishes
a weak first order transition from a second order one.
At the pseudo critical points, pseudo critical exponents
can be estimated. 
\end{abstract}


\maketitle

In recent years, much progress has been achieved in non-equilibrium
critical dynamics. For example, in a dynamic process in which 
a system initially at a high temperature or a zero temperature
state, is suddenly quenched to the critical temperature or nearby
and then evolves dynamically, {\it short-time} universal scaling 
behaviour has been found \cite {jan89,hus89}. 
This phenomenon is rather fundamental.
It exists not only in stochastic dynamics described
by Langevin equations \cite {jan89,oer93} 
or Monte Carlo algorithms \cite {hus89,hum91,sta92,li94,gra95,sch95},
but also in deterministic dynamics described by fundamental
microscopic equations of motion \cite {zhe99}.
More interestingly, based on the short-time scaling form,
it is possible to determine not only dynamic exponents
but also static exponents as well as the {\it critical} temperature
\cite {li95,luo98}.
Since the measurements are carried out in the short-time regime,
one does {\it not} suffer from critical slowing down.
Compared with {\it non-local} cluster algorithms, 
the short-time dynamic approach does study properties of the 
original local dynamics and also applies to systems with quenched
randomness. For a review, see Ref. \cite {zhe98}.

Naturally, it is interesting and attractive to
explore possible applications
of short-time dynamics to {\it first} order phase transitions.
Especially, due to large correlation lengths
and small discontinuities,
a {\it weak} first order transition presents
quite similar behaviour as a second order one. 
It has long been challenging how to distinguish
one from the other. Furthermore, {\it slowing down}
in Monte Carlo simulations at first order transitions
is even more severe than at second order ones.
Non-local cluster algorithms also do not show
much more efficiency.

In numerical simulations at first order transitions
{\it in equilibrium}, to locate the transition point
one usually searches for the maximums of the specific heat, 
susceptibility, or a Binder cumulant constructed from
energy \cite {bin92}. For a system with lattice size $L$,
these maximums deviate from the real transition point
by a power law $1/L^d$. To remove this power law
deviation, special techniques have been introduced
\cite {bor92}.
With these techniques, first order transition points 
can be determined rather accurately from moderate
lattice sizes, even for weak first order transitions.

To distinguish a first order transition
from a second order one, 
naively one may explore a signal for discontinuity of
the order parameter by increasing the lattice sizes.
Refined methods are typically based on
 the {\it finite size scaling}
of the specific heat, susceptibility, order parameter,
Binder cumulant of energy, or the transition point, e.g. see Refs.
\cite {bin92,pri90,bin97,fer88,fer89,oli95,jan95,fer98}.
However, when a first order transition is very weak,
it becomes subtle. The lattice sizes
one reaches in simulations hardly feel the difference between
very large correlation lengths in weak first order transitions
and divergent ones in second order transitions.
The double peak structure of the energy distribution
together with the finite size scaling
 shows its merit in this respect \cite {lee90,lee91,jan00},
but further efficient methods are still desired.

In this letter, we propose a short-time dynamic approach 
to weak first order transitions. The idea is inspired by
the existence of two {\it pseudo} critical points
$K^{*}$ and $K^{**}$  near
 the weak first order transition point $K_c$
 with $K^{**}<K_c<K^*$ \cite {gen75,fer92}.
In equilibrium, numerical measurements of $K^{*}$ and $K^{**}$
are not easy since they are induced by
{\it metastable} states. However, in short-time dynamics
$K^{*}$ and $K^{**}$ can be determined rather accurately
from two dynamic processes starting from
 high temperature and zero temperature states.
 In second order transitions, $K^{*}$ and $K^{**}$
 overlap with the transition point $K_c$.
 Therefore, difference of $K^{*}$ and $K^{**}$ 
 gives a criterion for
 a weak first order transition.
 
 As examples, we investigate the two-dimensional 
 $q$-state Potts models. The transition point is exactly
 known at $K_c=\ln(1+\sqrt q)$. 
 The phase transition is second order for $q\le4$
 and becomes first order for $q\ge5$.
 For small $q$, the first order transitions are weak.
 Especially, for $q=5$ the transition is so weak that 
 with standard methods one 
 hardly sees a difference from a second order one.

 In second order transitions, it has been shown that
 at the critical point, short-time behaviour
 of physical observables is a power law
 in dynamic processes starting from
  {\it both} a random and an ordered state.
 Away from the critical point, the power law behaviour is modified by
 a scaling function \cite {zhe98}.
 We will demonstrate that it is
  different for first order transitions.
  An approximate power law behaviour
  will be observed only at the pseudo critical points 
  $K^{*}$ and $K^{**}$ in proper dynamic processes.
 
 We begin our investigation by determining
$K^{*}$ for the 7-state Potts model. 
For this purpose, we consider a dynamic process
in which the system initially in a {\it random} state,
is suddenly quenched to $K_c$ or {\it above}, then evolves dynamically.
We have performed simulations with the heat-bath algorithm.
Lattice sizes are $L=140$ and $280$ and maximum updating times are
$t_{max}=2000$ and $6000$ respectively. Total samples for averaging
are $4600$ and errors are simply estimated by dividing
the data into four subsamples.

In Fig.~\ref{f1}(a), the second moment $M^{(2)}(t)$
 with $L=280$ is displayed for 
 $K=1.293562$ ($K_c$), $1.294210$ and $1.294857$
 on a log-log scale.
 Apparently, at $K_c$ the curve bends {\it downwards}
 and does not show a power law behaviour due to
 the random initial state and the
 {\it finite} spatial correlation length in equilibrium.
 Actually, this already indicates
 that the transition is first order
 if we assume $K_c$ is known.
 What is interesting here
  is that at a slightly bigger $K$, which we denote by
  $K^{*}$,
 one observes an approximate power law behaviour. 
 The {\it weaker} the transition is, 
 the {\it cleaner} the power law behaviour will be.
 When $K$ becomes bigger than $K^{*}$, the curve 
 bends {\it upwards}. Therefore, 
 $K^{*}$ looks like a critical point \cite {zhe98}.
 We can not prove that our
 $K^{*}$ is the same as the pseudo critical point $K^{*}$
 defined in equilibrium, but we strongly believe so.
 In equilibrium, $K^{*}$ is defined as a point
 at which the system presents approximate scaling
 behaviour similar to that at a critical point \cite {gen75,fer92}.
 
\begin{figure}[h]\centering  
\epsfysize=12cm 
\epsfclipoff 
\fboxsep=0pt
\setlength{\unitlength}{1cm} 
\begin{picture}(10,10)(0,0)
\put(0,8.5){{\epsffile{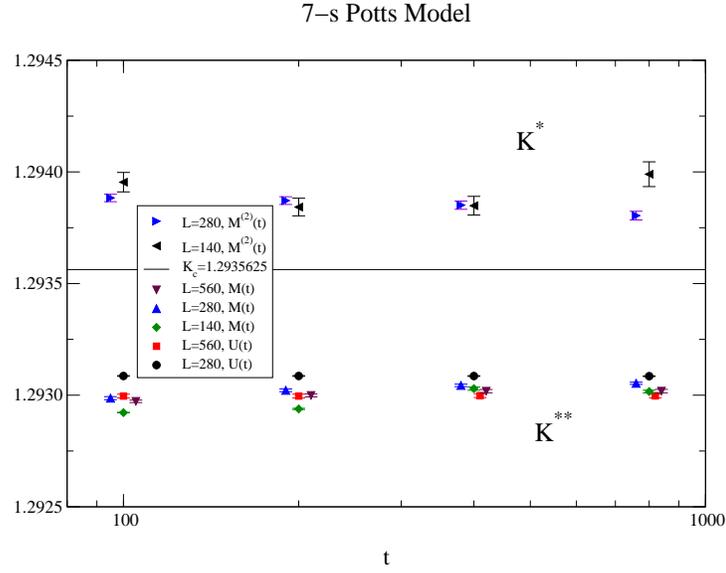}}} 
\end{picture} 
\caption{7-state Potts model.
The upper part shows the values for $K^*$ 
obtained from $M^{(2)}(t)$ in the interval $[t,t_{max}]$. 
The lower part shows $K^{**}$ 
obtained from $M(t)$ and $U(t)$. 
The line denotes the exact value $K_c$.
} 
\label{f2}
\end{figure}

 In our short-time dynamic approach, 
 practically we locate
 the pseudo critical point $K^{*}$
  by interpolating $M^{(2)}(t)$  among
 the three simulated $K$'s and searching for 
 the best power law behaviour \cite {luo98,zhe98}.
 In short-time critical dynamics,
 it has been intensively discussed that 
 universal behaviour emerges only after a time scale 
 $t_{mic}$ which is large enough in microscopic sense.
 If a Monte Carlo time step (a sweep over all spins
 on the lattice) is considered to be
 a microscopic time unit, $t_{mic}$ is typically
 $10$ to some hundred time steps \cite {zhe98}.
 Similarly, in first order transitions,
 physical behaviour at {\it macroscopic} level
 is presented also only after $t_{mic}$.
 In the upper part of Fig.~\ref{f2}, 
 $K^*$ obtained with data in a time interval 
$[t,t_{max}]$ is shown. The results are stable
and $K^*$ is clearly above $K_c$.
The final value for $K^*$ is estimated to be  
 $K^{*}_{7s} = 1.293854(29)$. 
 This is consistent with the value
 $K^{*}=1.2945(9)$ given in Ref.~\cite{fer92}.
 However, the latter can hardly distinguish
 $K^*$ from $K_c$ within the error.
 
 To determine $K^{**}$, we study a dynamic process
in which the system initially in an {\it ordered} state,
is quenched to $K_c$ or {\it below}, and evolves dynamically.
Here we have performed extra simulations for $L=560$,
up to $t_{max}=6000$.
Total samples for $L=140$, $280$ and $560$ are
$7000$, $1500$ and $135$ respectively.
In Fig.~\ref{f1}(b) the magnetisation with $L=280$
is plotted for $K=1.2929$, $1.2930$ and $1.2931$.
The curve for $K_c=1.293562$ (not in the figure)
is much above that for $1.2931$
 and very far from power law behaviour. This is again a signal
 for a first order transition.
 The reason is clear. For first order transitions,
 with an ordered initial state the system 
 will evolve to the ordered phase at $K_c$.
 However, at the pseudo critical point $K^{**}$
 we will observe approximate
 power law behaviour.
 Searching for a curve with 
 the best power law behaviour
 from the three curves in Fig.~\ref{f1}(b),
 we determine the pseudo critical point $K^{**}$.
The results are presented in 
the lower part of Fig.~\ref{f2}. The values
are clearly below $K_c$.

Another interesting observable is the
Binder cumulant $U(t) \equiv M^{(2)}(t)/(M(t))^2-1$.
If a transition is second order,
$U(t)$ obeys
a power law at the transition point.
  Therefore it can also be used for
the determination of $K^{**}$.
Results are included in Fig.~\ref{f2}.
Summarising all these measurements leads to 
$K^{**}_{7s} = 1.293008(7)$.

\begin{figure}[h]\centering  
\epsfysize=12cm 
\epsfclipoff 
\fboxsep=0pt
\setlength{\unitlength}{1cm} 
\begin{picture}(9,9)(0,0)
\put(0,8.5){{\epsffile{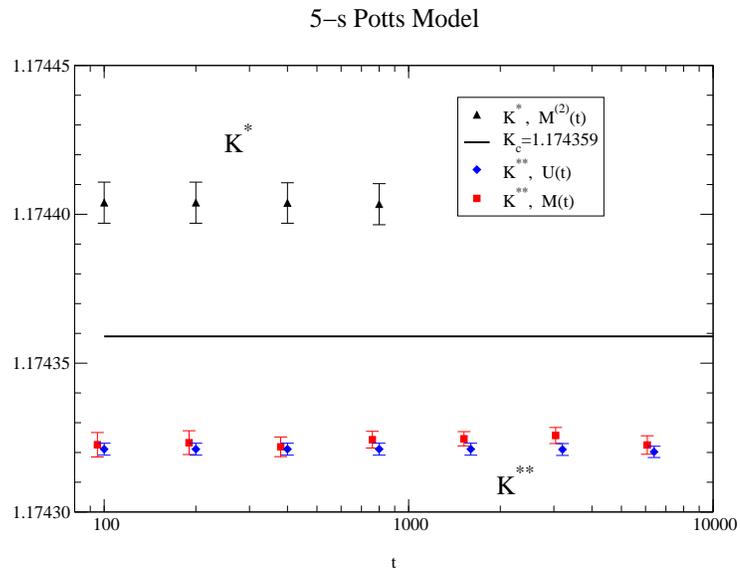}}} 
\end{picture} 
\caption{
5-state Potts model.
The upper part shows the values for $K^*$ 
obtained from $M^{(2)}(t)$ with $L=560$ in the interval 
 $[t,t_{max}=10000]$. 
The lower part shows $K^{**}$
obtained from $M(t)$ with $L=560$ in the interval 
 $[t,t_{max}=40000]$.
The line denotes the exact value $K_c$.
} 
\label{f4}
\end{figure}

For the 5-state Potts model, the transition is extremely weak.
One should carry out the simulations very carefully.
To locate $K^{*}$, we have first performed simulations
with $L=560$ for $K=1.174359$ ($K_c$), $1.174946$, and $1.175533$,
up to $t_{max}=10\;000$ with 
$1800$ samples. The resulting $K^*_{5s}=1.17445(6)$ 
is not accurate enough. Therefore another simulation has been carried out
at $K=1.174570$, which is much closer to $K^{*}$.
In Fig.~\ref{f3}(a), the second moments for $K=1.174359$ ($K_c$)
and $1.174570$ are displayed.
With these data more accurate values for $K^{*}$
 are obtained and collected
in the upper part of Fig.~\ref{f4}. We estimate the
averaged $K^*_{5s}=1.174404(7)$.

Similar is the case for the determination of $K^{**}$.
We have first performed simulations with an
ordered initial state with lattice sizes 
$L=280$ and $560$ for 
$K=1.173890$, $1.174125$, and $1.174359$ ($K_c$), 
up to $t_{max}=10\;000$ with total samples $725$.  
From the data for the magnetisation we estimate 
a relatively rough $K^{**}_{5s}=1.17428(9)$. Then we performed
simulations at $K=1.174280$ and $1.174359$ ($K_c$)
up to $t_{max}=40\;000$. 
The results are not sensitive to whether we take
$t_{max}=10\;000$ or $40\;000$.
In Fig.~\ref{f3}(b), the magnetisation at 
$K=1.174280$ and $1.174359$ ($K_c$) are plotted.
From the lower part of Fig.~\ref{f4}, we obtain 
a final value $K^{**}_{5s}=1.174322(2)$.
 
In Table~\ref{ta1}, all results for $K^*$ and $K^{**}$ 
have been collected. 
For both the 7-state and the 5-state Potts model,
 $K^{*}$ and $K^{**}$ are clearly  
 above and below the transition point $K_c$ respectively. 
 Our short-time dynamic approach indeed provides a safe 
criterion for a weak first order transition.

Since our dynamic measurements are carried out in the short-time regime
when the spatial correlation length is still short,
we can easily control the finite size effect.
We also do not have the problem of generating independent
configurations and therefore do not suffer from 
slowing down. After excluding the finite size effect,
the measurements are sensitive enough to distinguish
a finite but very large spatial correlation length in equilibrium
from an infinite one.  This is why our method is
successful.

With the pseudo critical points in hand,
assuming similar scaling laws as in
second order transitions \cite {zhe98}, 
one can estimate corresponding
pseudo critical exponents. At $K^{*}$, e.g,
\begin{equation}
M^{(2)}(t) \propto t^{c_2}\ , \quad 
         c_2=(d-2\beta/\nu)/z.
\label{fo4}
\end{equation}
At $K^{**}$, for the magnetisation,
\begin{equation}
M(t)\propto t^{-c_1}\ , \quad c_1=\beta/\nu z,
\label{fo1}
\end{equation}
while for the Binder cumulant 
\begin{equation}
U(t)\propto t^{c_U}\ , \quad c_U=d/z.
\label{fo2}
\end{equation}
Here $d$ is the dimension of the lattice,  $\beta$
and $\nu$ are the well known static exponents
and $z$ is the dynamic exponent. 
However, the values of the exponents at $K^{*}$ and $K^{**}$ 
can be different. Plotting the observables vs. $t$
in log-log scale, one measures the corresponding exponents from
the slopes. The results are given in Table~\ref{ta2}.
Here we should admit that accurate values for 
complete sets of exponents can not be obtained
so easily. One still needs much more careful simulations.
 An important reason is that
$K^{*}$ and $K^{**}$ are not real critical points.
They are also rather close to each other.

In conclusions, we have proposed a short-time dynamic
approach to weak first order transitions.
From non-equilibrium short-time behaviour
of two dynamic processes starting from
random and ordered initial states,
pseudo critical points $K^{*}$ and $K^{**}$
are determined. Difference of $K^{*}$ and $K^{**}$
distinguishes a weak first order transition from
a second order one. Since the measurements are carried out
in short-time regimes, the method
does not suffer from slowing down.
Different from many techniques developed
in simulations in equilibrium,
our method is not based on the finite size scaling.

A simple average of $K^{*}$ and $K^{**}$ gives a rather good
estimate of the transition point $K_c$,
especially for very weak transitions.
For example, for the 5-state Potts model
$(K^{*}+K^{**})/2=1.174363$ and the relative deviation from 
the exact $K_c$ is only the order of $O(10^{-6})$.
It is interesting to investigate
how to obtain an accurate $K_c$ for not too weak transitions.
Furthermore, how other relevant observables like
 the specific heat and energy distribution
evolve in non-equilibrium dynamics is also
an important topic. It is challenging whether
from short-time dynamics one can estimate the latent heat
and the discontinuity of the order parameter
in equilibrium.

{\bf Acknowledgements}:
The authors thank Z.B. Li and Q. Wang for valuable comments.
This work is supported in part by the Deutsche Forschungsgemeinschaft, 
Az. Schu 95/9-2 and TR 300/3-1.


\begin{figure}[h] 
\epsfysize=8.5cm 
\epsfclipoff 
\fboxsep=0pt
\setlength{\unitlength}{1cm} 
\begin{picture}(9,6.5)(0,0)
\put(-.5,6.2){{\epsffile{M2_7s_280.ps}}} 
\epsfysize=8.5cm 
\put(7.5,6.2){{\epsffile{M_t_280.ps}}} 
\put(0.7,0.3){(a)}
\put(9.3,0.3){(b)}
\end{picture} 
\caption{
7-state Potts model:
(a) Second moment $M^{(2)}(t)$ plotted vs. $t$ 
on log-log scale for $K=1.293562$ ($K_c$), $1.294210$
 and $1.294857$ (from below) with $L=280$.
(b) Magnetization $M(t)$ plotted vs. $t$ on log-log scale
for $K=1.2929$,
$1.2930$ and $1.2931$ (from below) with $L=280$.
} 
\label{f1}
\end{figure}

\begin{figure}[h] 
\epsfysize=8.5cm 
\epsfclipoff 
\fboxsep=0pt
\setlength{\unitlength}{1cm} 
\begin{picture}(9,4.5)(0,0)
\put(-.5,6.2){\epsffile{M2l_560.ps}} 
\epsfysize=8.5cm 
\put(7.5,6.2){\epsffile{M1l_560.ps}} 
\put(0.7,0.3){(a)}
\put(8.3,0.3){(b)}
\end{picture} 
\caption{
5-state Potts model:
(a) $M^{(2)}(t)$ plotted vs. $t$ on log-log scale for 
 $K_c=1.174359$ and $K=1.174570$ with $L=560$.
(b) $M(t)$ plotted vs. $t$ on log-log scale
for $K=1.174280$ and $K_c=1.174359$ with $L=560$.
} 
\label{f3}
\end{figure}

\begin{table}[h]\centering 
\begin{tabular}{l|l|l|l}
 & $K^{**}$ & $K_c$ & $K^{*}$ \\ 
 \hline 
 $q=5$ & 1.174322(2) &  1.174359 & 1.174404(7) \\ 
\hline
$q=7$&  1.293008(7) & 1.293562 &  1.293854(29) 
\end{tabular} 
\caption{ Pseudo critical points $K^{**}$ and $K^*$ 
measured from short-time dynamics 
for the 5-state and 7-state Potts models, in comparison with
 the transition point $K_c$. } 
\label{ta1} 
\end{table} 

\begin{table}[h]\centering
\begin{tabular}{l|l|l||l}
& ${c^{**}_1}$ & ${c^{**}_U}$ &
  $c^{*}_2$  \\
\hline
$q = 5 $ & 0.091(2) & 0.93(3) &   0.716(3)  \\
\hline
$ q = 7$  & 0.0239(2) & 0.885(8) & 0.502(5)  \\
\hline
\end{tabular} 
\caption{
Pseudo critical exponents for the 5-state and 7-state Potts models.
}
\label{ta2}
\end{table}

\end{document}